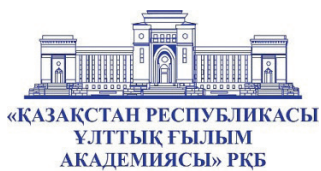
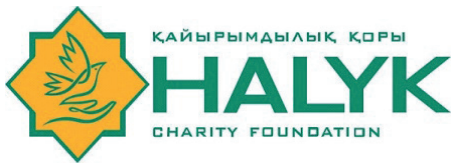

«ҚАЗАҚСТАН РЕСПУБЛИКАСЫ
ҰЛТТЫҚ ҒЫЛЫМ АКАДЕМИЯСЫ» РҚБ
«ХАЛЫҚ» ЖҚ

# БАЯНДАМАЛАРЫ

# ДОКЛАДЫ
РОО «НАЦИОНАЛЬНОЙ
АКАДЕМИИ НАУК РЕСПУБЛИКИ КАЗАХСТАН»
ЧФ «ХАЛЫҚ»

# REPORTS
OF THE ACADEMY OF SCIENCES
OF THE REPUBLIC OF KAZAKHSTAN
«Halyk» Private Foundation

PUBLISHED SINCE JANUARY 1944

ALMATY, NAS RK

© **S.B. Dubovichenko[1], N.A. Burkova[2], A.S. Tkachenko[1], D.M. Zazulin[2*]**, 2023
[1] Fesenkov Astrophysical Institute, Almaty, Kazakhstan;
[2] Al-Farabi Kazakh National University, Almaty, Kazakhstan.
*E-mail: denis_zazulin@mail.ru

# REACTION RATE OF RADIATIVE CAPTURE PROTON BY $^{10}$B

**Dubovichenko Sergey Borisovich** - Laureate of the al-Farabi State Prize of the Republic of Kazakhstan in the field of science and technology, doctor of physical and mathematical sciences, professor, head of laboratory of V.G. Fesenkov Astrophysical Institute, Almaty, Kazakhstan
E-mail: dubovichenko@gmail.com, http://orcid.org/0000-0002-7747-3426;
**Burkova Nataliya Aleksandrovna** - doctor of physical and mathematical sciences, professor of department of theoretical and nuclear physics of Al-Farabi Kazakh National University, Almaty, Kazakhstan
E-mail: natali.burkova@gmail.com, https://orcid.org/0000-0002-3122-1944;
**Tkachenko Alessya** - Ph.D., researcher of V.G. Fesenkov Astrophysical Institute, Almaty, Kazakhstan,
E-mail: tkachenko.alessya@gmail.com, https://orcid.org/0000-0002-9319-0135;
**Zazulin Denis Mikhailovich** - candidate of physical and mathematical sciences, acting associate professor of al-Farabi Kazakh National University, Almaty, Kazakhsta
E-mail: denis_zazulin@mail.ru, https://orcid.org/0000-0003-2115-6226.

**Abstract.** The $^{10}$B(p,γ)$^{11}$C reaction is of significant interest in nuclear astrophysics and in the field of controlled thermonuclear fusion. This reaction is one of the reactions of $^{11}$B production, which is carried out through the $^{10}$B(p,γ)$^{11}$C(β+ν)$^{11}$B chain. The rate of the $^{10}$B(p,γ)$^{11}$C reaction (occurring in the interiors of first-generation stars) can be of great importance for the amount of $^{10}$B and $^{11}$B observed today in the interstellar medium and in the Earth's crust. In thermonuclear reactors, structural elements containing boron can be used as neutron absorbers, etc. Therefore, in this work, within the framework of a modified potential cluster model with a classification of orbital states according to Young's diagrams and taking into account allowed and forbidden states, we examined the possibility of describing the available experimental data for the total cross sections of the radiative $p^{10}$B capture to the ground state of the $^{11}$C nucleus at energies up to 1 MeV. It is shown that only on the basis of $E$1 and $M$1 transitions from the $p^{10}$B scattering states, taking into account the first resonance for the ground state of the $^{11}$C nucleus, it is quite possible to explain the magnitude and shape of the experimental astrophysical S-factor. The work presents comparisons the astrophysical S-factors of the radiative $p^{10}$B capture to the ground state of the $^{11}$C nucleus found by us with the experimental data available in the literature. Based on the obtained theoretical S-factor, the rate of this reaction was calculated in the temperature range from 0.01 to 1 $T_9$. The calculated results for rates are approximated by a simple expression, which simplifies their use in applied thermonuclear and astrophysical research.

© С.Б. Дубовиченко[1], Н.А. Буркова[2], А.С. Ткаченко[1], Д.М. Зазулин[2*], 2023
[1]В.Г. Фесенков Атындағы Астрофизика Институты, Алматы, Қазақстан;
[2]Әл-Фараби Атындағы Қазақ Ұлттық Университеті, Алматы, Қазақстан.
*E-mail: denis_zazulin@mail.ru

## $^{10}$B РАДИЯЛЫҚ ПРОТОНДЫ ТҮСІРУ ҚАРҚЫМЫ

**Дубовиченко Сергей Борисович** – ғылым және техника саласындағы әл-Фараби атындағы Қазақстан Республикасының Мемлекеттік сыйлығының Лауреаты, физика-математика ғылымдарының докторы, профессор, атындағы Астрофизика институтының зертхана меңгерушісі В.Г. Фесенкова, Алматы, Қазақстан
E-mail: dubovichenko@ gmail.com, http://orcid.org/0000-0002-7747-3426;

**Буркова Наталья Александровна** – физика-математика ғылымдарының докторы, әл-Фараби атындағы Қазақ ұлттық университетінің теориялық және ядролық физика кафедрасының профессоры, Алматы, Қазақстан
E-mail: natali.burkova@gmail.com,https://orcid. org/0000-0002-3122-1944;

**Ткаченко Алеся** - Ph.D., атындағы Астрофизика институтының ғылыми қызметкері В.Г. Фесенкова, Алматы, Қазақстан
E-mail: tkachenko.alessya@gmail.com, https://orcid.org/0000-0002-9319-0135;

**Зазулин Денис Михайлович** – физика-математика ғылымдарының кандидаты, әл-Фараби атындағы Қазақ ұлттық университетінің теориялық және ядролық физика кафедрасының доцентінің, Алматы, Қазақстан
E-mail: denis_zazulin@mail.ru, https://orcid.org/0000-0003-2115-6226.

**Аннотация**. $^{10}$B$(p,\gamma)^{11}$C реакциясы ядролық астрофизикада және басқарылатын термоядролық синтез саласында маңызды қызығушылық тудырады. Бұл реакция $^{10}$B$(p,\gamma)^{11}$C$(\beta+\nu)^{11}$B тізбегі арқылы жүзеге асырылатын $^{11}$B өндірісі реакцияларының бірі. $^{10}$B$(p,\gamma)^{11}$C реакциясының жылдамдығы (бірінші ұрпақ жұлдыздарының ішкі қабаттарында болатын) қазіргі таңда жұлдыз аралық ортада және жер қыртысында байқалатын $^{10}$B және $^{11}$B мөлшері үшін үлкен маңызға ие болуы мүмкін. Термоядролық реакторларда құрамында боры бар құрылымдық элементтер нейтронды сіңіргіштер және т.б. ретінде қолданылады. Сондықтан, осы жұмыста біз Юнг схемалары бойынша орбиталық күйлерді жіктеумен өзгертілген кластерлік модель шеңберінле және рұқсат етілген және тыйым салынған күйлерді ескере отырып, 1 МэВ дейінгі энергиялар кезінде $^{11}$C ядросының негізгі күйіне түсірудің радиациялық $p^{10}$B толық қималары үшін қолда бар эксперименттік деректерді сипаттау мүмкіндігін қарастырдық. Тек $p^{10}$B шашырау күйлерінен $E$1 және $M$1 ауысуларының негізінде $^{11}$C ядросының негізгі күйіне бірінші резонансты ескере отырып, тәжірибелік астрофизикалық S - факторының шамасы мен пішінін түсіндіруге әбден болатыны көрсетілген. Бұл мақалада біз тапқан $^{11}$C ядросының негізгі күйіне $p^{10}$B сәулеленуінің астрофизикалық S-факторлары әдебиеттегі эксперименттік дерктермен салыстырылды. Алынған теориялық S-фактор негізінде бұл реакцияның жылдамдығы 0.01-ден 1 $T_9$-ға дейінгі температура диапазонында есептелді. Жылдамдықтар үшін есептелген нәтижелер оларды қолданбалы термоядролық және астрофизикалық зерттеулерде қолдануды жеңілдететін қарапайым өрнекпен жуықталады.

**Түйін сөздер:** Ядролық астрофизика, жеңіл атомдық ядролар, төмен және астрофизикалық энергиялар, радиацияны түсіру, термоядролық процестер, потенциалды кластерлік модель, Юнг схемасы

© **С.Б. Дубовиченко[1], Н.А. Буркова[2], А.С. Ткаченко[1], Д.М. Зазулин[2\*]**, 2023

[1]Астрофизический Институт Имени В.Г. Фесенкова, Алматы, Казахстан;
[2]Казахский Национальный Университет Имени Аль-Фараби,
Алматы, Казахстан.
E-mail: denis_zazulin@mail.ru

# СКОРОСТЬ РАДИАЦИОННОГО ЗАХВАТА ПРОТОНОВ НА $^{10}$B

**Дубовиченко Сергей Борисович** – Лауреат Государственной премии Республики Казахстан имени аль-Фараби в области науки и техники, доктор физико-математических наук, профессор, заведующий лабораторией Астрофизического Института им. В.Г. Фесенкова, Алматы, Казахстан
E-mail: dubovichenko@gmail.com, ttp://orcid.org/0000-0002-7747-3426;
**Буркова Наталья Александровна** - доктор физико-математических наук, профессор кафедры теоретической и ядерной физики Казахского Национального Университета им. аль-Фараби, Алматы, Казахстан
E-mail: natali.burkova@gmail.com, https://orcid.org/0000-0002-3122-1944;
**Ткаченко Алеся** - Ph.D., научный сотрудник Астрофизического Института им. В.Г. Фесенкова, Алматы, Казахстан
E-mail: tkachenko.alessya@gmail.com, https://orcid.org/0000-0002-9319-0135;
**Зазулин Денис Михайлович** – кандидат физико-математических наук, и.о. ассоциированного профессора Казахского Национального Университета им. аль-Фараби, Алматы, Казахстан
E-mail: denis_zazulin@mail.ru, https://orcid.org/0000-0003-2115-6226.

**Аннотация.** Реакция $^{10}$B$(p,\gamma)^{11}$C представляет существенный интерес в ядерной астрофизике и в области управляемого термоядерного синтеза. Эта реакция является одной из реакций производства $^{11}$B, которое осуществляется через цепочку $^{10}$B$(p,\gamma)^{11}$C$(\beta^+\nu)^{11}$B. Скорость реакции $^{10}$B$(p,\gamma)^{11}$C (протекавшей в недрах звезд первого поколения) может иметь большое значение для наблюдаемого сегодня количества $^{10}$B и $^{11}$B в межзвездной среде и в земной коре. В термоядерных реакторах конструкционные элементы, содержащие бор могут использоваться в качестве поглотителей нейтронов и т.д. Поэтому нами, в данной работе, в рамках модифицированной потенциальной кластерной модели с классификацией орбитальных состояний по схемам Юнга и с учетом разрешенных и запрещенных состояний рассмотрена возможность описания имеющихся экспериментальных данных для полных сечений радиационного $p^{10}$B захвата на основное состояние ядра $^{11}$C при энергиях до 1 МэВ. Показано, что только на основе $E$1- и $M$1-переходов из состояний $p^{10}$B рассеяния с учетом первого резонанса на основное состояние ядра $^{11}$C вполне удается объяснить величину и форму экспериментального астрофизического $S$-фактора. В работе приведены сравнения найденных нами астрофизических S-факторов радиационного $p^{10}$B захвата на основное состояние ядра $^{11}$C с имеющимися в литературе экспериментальными данными. На основе полученного теоретического $S$-фактора рассчитана скорость этой реакции в области температур от 0.01 до 1 $T_9$. Расчетные результаты для скоростей аппроксимируются простым выражением, что упрощает их использование в прикладных термоядерных и астрофизических исследованиях.

**Ключевые слова:** Ядерная астрофизика, легкие атомные ядра, низкие и астрофизические энергии, радиационный захват, термоядерные процессы, потенциальная кластерная модель, схемы Юнга

## Introduction

Here are the results in the field of research of the thermonuclear capture reaction $p^{10}$B at low and astrophysical energies. This reaction is not directly included in the thermonuclear cycles, and so far, apparently, it has been considered in detail only in our work (Dubovichenko, 2015 a). In it, as in the work of (Burkova, 2021), as a nuclear model, we used a modified potential cluster model (MPCM), which allows us to consider some thermonuclear processes, namely, reactions of radiative capture of nucleons and the lightest clusters by light nuclei, based on unified concepts, criteria and methods (Dubovichenko, 2015 b). The model takes into account the classification of states according to Young's diagrams, which makes it possible to determine the presence of forbidden states (FS) and allowed states (AS) in intercluster potentials.

Previously, in our work (Dubovichenko, 2015 a), for the capture of $p^{10}$B to the ground state (GS), we obtained the astrophysical S-factor at energies up to 1 MeV, which generally described the available experimental data. Here we also consider the S-factor up to 1 MeV, but we perform the refinement of potentials with forbidden states and determine the rate of this reaction in the range from 0.01 to 1 $T_9$.

**Model and calculation methods. Structure of levels of the $p^{10}$B system.** The bound allowed $p^{10}$B state in the $^6P_{3/2}^{g.s.}$ - wave corresponds to the GS of $^{11}$C with $J^\pi,T = 3/2^-,1/2$ and the Young's diagram of {443} (Dubovichenko, 2015 a) and is at the binding energy of -8.6894 MeV of the $p^{10}$B system (Kelley, 2012) (recall that for $^{10}$B $J^\pi,T = 3^+,0$ is known (Kelley, 2012)). Some $p^{10}$B scattering states and BS can be mixed in spin with $S = 5/2$ (2$S$+1 = 6) and $S = 7/2$ (2$S$+1 = 8), but since we consider only transitions to $^6P_{3/2}^{g.s.}$ of GS, in what follows calculations only partial waves with spin $S = 5/2$ will be used.

Let us now consider the spectrum of resonant levels of the $^{11}$C nucleus in the $p^{10}$B channel at energies below 1.0 MeV, it has the following states (Kelley, 2012):

1. Resonance at energy of 10(2) keV in the c.m. with angular momentum of $J = 5/2^+$ and width 15(1) keV in c.m. It corresponds to the level of the $^{11}$C nucleus at excitation energy of 8.699(2) MeV (Kelley, 2012) and can be associated with the $^6S_{5/2}^1$ or $^6D_{5/2}$ states.

2. State at energy of 511(50) keV in the c.m. with angular momentum of $J = 5/2^+$ and width 500(90) keV in c.m. (Kelley, 2012) is the $^{11}$C level at energy of 9.200(50) MeV, which can also be associated with the $^6S_{5/2}^1$ or $^6D_{5/2}$ states.

3. Third resonance at 0.941(50) MeV in the c.m. with angular momentum of $J = 5/2^-$ and width 271(60) keV in c.m. is the level of the $^{11}$C nucleus at energy of 9.630(50) MeV (Kelley, 2012), which can be associated with the $^6P_{5/2}^1$ state. However, its moment is not precisely defined, and we will not consider it.

4. Fourth resonance at 0.956(50) MeV in the c.m. with angular momentum $J = 3/2^-$ and width 378(56) keV in c.m. is the level of the $^{11}$C nucleus at energy of 9.645(50) MeV (Kelley, 2012), which can be attributed to the $^6P_{3/2}$ state, but its angular momentum is not precisely determined, and we will not consider it.

5. The next resonance at 9.780(30) or 1.091(30) MeV above the threshold with an inaccurate determined angular momentum of $J = 5/2^-$ has a width of 240(50) keV (Kelley, 2012) and can be attributed to the $^6P_{5/2}^1$ state. We also will not consider it because of an uncertain angular momentum.

6. At higher energies, several resonances with different angular momentums are observed (Kelley, 2012), which either belong to the spin channel with $S = 7/2$ or to the $F$-wave and will not be considered by us.

At low energies, transitions are possible mainly from the $S$-wave of scattering; therefore, when considering $E1$ transitions, they are possible only to the $P$-bound state. Note that based on the shape of the S-factor of $p^{10}$B capture, the resonance at 0.511 MeV in the $^6S_{5/2}^1$ - wave is practically not noticeable due to its large width. Therefore, the $E1$ transition from the $^6S_{5/2}^1$ - scattering wave to the $^6P_{3/2}^{g.s.}$ of GS ($^6S_{5/2}^1 \to {}^6P_{3/2}^{g.s.}$) is possible. $M1$-transition from $P$-scattering waves to the GS is also possible. If we assume that a potential different from $^6P_{3/2}^{g.s.}$ of GS is used for the $^6P_{3/2}$ - wave, two transitions are possible ($^6P_{3/2} + {}^6P_{5/2} \to {}^6P_{3/2}^{g.s.}$).

**Construction of potentials of the $p^{10}$B interactions.** To construct central intercluster potentials, we use the Gaussian type of interaction

$$V(r,JLS) = -V_0(JLS)\exp\{-\gamma(JLS)r^2\}$$

with a point Coulomb term. Since, in what follows, transitions from some scattering waves to the $^6P_{3/2}^{g.s.}$ of GS of the $^{11}$C nucleus in the $p^{10}$B channel will be considered first, we will obtain first the potential of this state. As already mentioned, the ground state is at a binding energy of -8.6894 MeV in the $p^{10}$B channel and has angular momentum of 3/2⁻, being pure in spin with the $S = 5/2$ of $^6P_{3/2}^{g.s.}$ level (Kelley, 2012).

Since we were unable to find data on the $^{11}$C charge radius, we assume that it differs little from the $^{11}$B radius, which is 2.43(11) fm (Kelley, 2012). The $^{10}$B radius is known and is equal to 2.4277(499) fm (http://cdfe.sinp.msu.ru), and the proton radius is 0.8775(51) fm (https://physics.nist.gov/cgi-bin/cuu /Value?mud%20 csearch_for=atomnuc!). In all calculations given below for the proton mass, the value of $m_p$ = 1.00727646677 a.m.u. was used. (https://physics.nist.gov/cgi-bin/cuu/ Value?mud%20csearch_for=atomnuc!), and the mass of $^{10}$B, $m(^{10}$B$)$ is 10.012936 amu. (http://cdfe.sinp.msu.ru).

As a result, the parameters of the $^6P_{3/2}^{g.s.}$ of GS potential without FS were obtained, as follows from the classification of FS and AS according to Young's diagrams given in (Dubovichenko, 2015a).

$$V_{g.s.} = 337.1459 \text{ МэВ}, \quad \alpha_{g.s.} = 1.0 \text{ Фм}^{-2}. \tag{1}$$

They lead to a charge radius of 2.32 fm, a binding energy of -8.6894 MeV with a relative accuracy of $10^{-4}$ MeV, and a dimensionless asymptotic constant (AC) of 1.16 in the range of 5 – 10 fm. The scattering phase of such a potential smoothly decreases from 180⁰ at zero energy to 179⁰ at 1.0 MeV.

As usual (Dubovichenko, 2015 b), we use the dimensionless quantity of AC, which is determined in terms of the Whittaker functions in the form $\chi_L(r) = \sqrt{2k_0}\, C_W W_{-\eta L + 1/2}(2k_0 r)$ (Plattner, 1981; Whittaker, 1903). In the work (Artemov, 2010) for the squared asymptotic normalization coefficient (ANC) $A_{NC}$ of GS, the value of 8.9(8) fm$^{-1}$ is given, which contains the factor "6" associated with the permutation of nucleons (Blokhintsev, 1977; Mukhamedzhanov, 1999). Then for the dimensional AC in $\chi_L(r) = CW_{-\eta L+1/2}(2k_0 r)$ we get the value of 1.22(5) fm$^{-1/2}$. The dimensionless AC used is related to the dimensional one by the formula of $C = \sqrt{2k_0}\, C_W$. For the relationship between AC and ANC, the expression of $A_{NC} = \sqrt{S_f}\, C$ is known; therefore, at $S_f = 1$ and $\sqrt{2k_0} = 1.11$, we have 1.10(5) for the dimensionless value of AC, which is in good agreement with the value obtained above.

Next, one can construct the potential for the $^6S_{5/2}$ resonance, which is at 0.0096(20) MeV with angular momentum of $J = 5/2^+$ and width of 15(1) keV and corresponds to the 8.699(2) MeV level of the $^{11}$C nucleus (Kelley, 2012). Such a potential may have parameters

$$V_{S52} = 49.95 \text{ МэВ}, \quad \alpha_{S52} = 0.088 \text{ Фм}^{-2}. \tag{2}$$

It contains FS (Dubovichenko, 2015a) and, as will be seen below, it almost correctly reproduces the behavior of the total cross sections for radiative $p^{10}$B capture (Wiescher, 1983; Tonchev, 2003) at the lowest energies. The parameters of such a potential were chosen solely for the correct description of the first resonance in the astrophysical S-factor of the $p^{10}B$ capture. We also assume that in the energy range that we are considering, $^6P_{3/2}$ and $^6P_{5/2}$ - scattering waves do not have resonances, so their phases must have a non-resonant character.

**Results and discations. Total cross sections for radiative $p^{10}B$ capture.** Experimental data for the total cross sections and astrophysical S-factors of the radiative $p^{10}B$ capture are given in (Wiescher, 1983; Tonchev, 2003; Hunt, 1957). Let us now consider the results of calculation of the astrophysical S-factor of the radiative $p^{10}B$ capture on the GS of the $^{11}C$ nucleus with potentials (1) and (2). The astrophysical S-factor does not contain explicit resonances in the energy range up to 1 MeV, as can be seen in Fig. 1. There is only a resonance in the zero energy region, which corresponds to a resonance in the $^6S_{5/2}$ scattering wave at 10(2) keV. As a result, the form of the calculated S-factor of the $p^{10}B$ capture was obtained for the $E1$ transition from $^6S_{5/2}$ - scattering wave to $^6P_{3/2}$ of GS (process that is shown in Fig. 1 by a dashed curve).

As can be seen from Fig. 1, the calculated S-factor adequately reproduces the results of experimental measurements from (Wiescher, 1983; Tonchev, 2003; Hunt, 1957) in the region of the first resonance and up to energy of approximately 0.25 MeV. Since the experimental S-factor above 300 keV is not of a resonant nature, we further considered only non-resonant $M1$ transitions to the $^6P_{3/2}$ of GS (process that is shown in Fig. 1 by a dotted curve). The continuous line in Fig. 1 shows the summary cross section of the two processes considered above, which describes the S-factor in the energy range from 50 keV to 1 MeV.

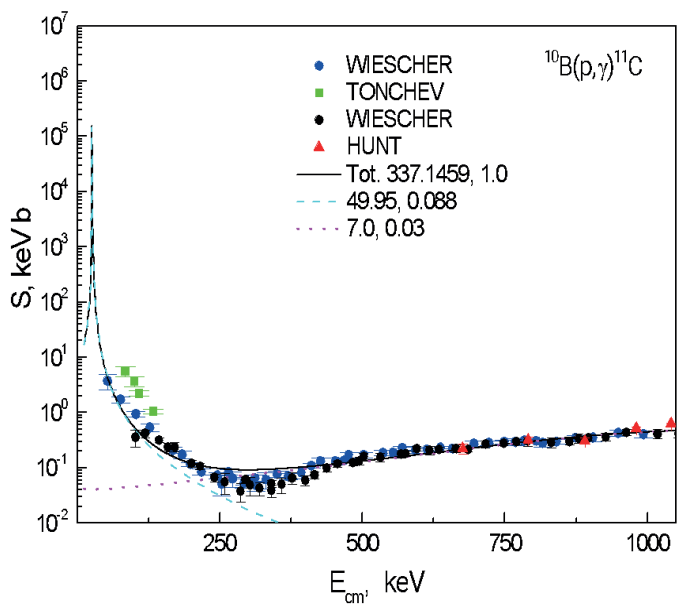

Picture 1. Astrophysical S-factor of radiative $p^{10}B$ capture on the GS. Dots and squares – experiments data (Wiescher, 1983; Tonchev, 2003; Hunt, 1957).

Since the results of phase analysis for the elastic $p^{10}B$ scattering are not available, and there are no exact phase values, it can be assumed that the phases of $^6P$- scattering in the energy range below 1.0 MeV do not have to be exactly equal to zero. Therefore, the parameters of the $P$-potentials for nonresonant $M1$ transitions were chosen so as to correctly convey the overall behavior of the S-factor $p^{10}B$ of capture at energies above 0.25 MeV.

It turned out that the potential parameters for both $^6P$- scattering waves without coupled FSs can be represented in the following form

$$V_P = 7 \text{ МэВ}, \quad \alpha_P = 0.03 \text{ Фм}^{-2} \tag{3}$$

They make it possible, on the whole, to correctly describe the available experimental data on the S-factor at energies from 0.25 to 1.0 MeV, as shown in Fig. 1 by the dotted curve.

**Reaction rate of the proton capture on $^{10}B$.** Further, in Fig. 2, the continuous curve shows the radiative capture rate of $p^{10}B$, which corresponds to the calculated astrophysical S-factor shown in Fig. 1.

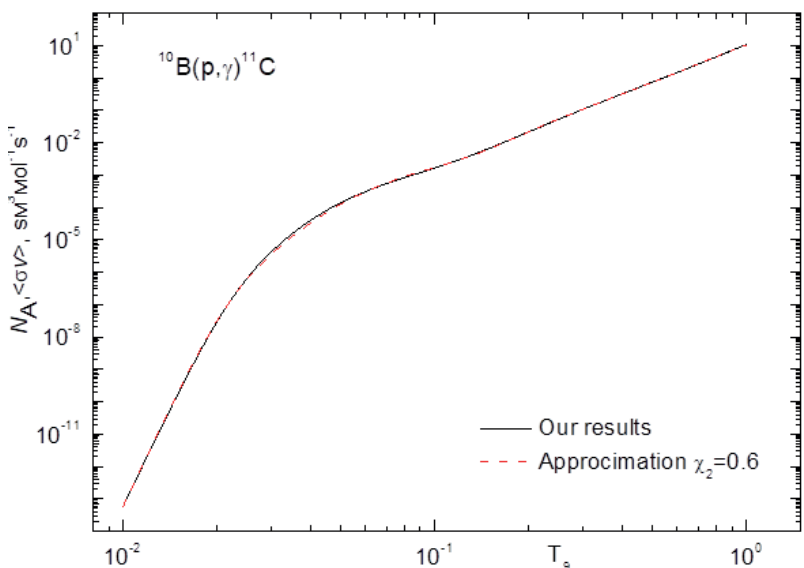

Figure 2. Rate of radiative $p^{10}B$ capture on the GS.

The calculated reaction rate shown in Fig. 2 as a continuous curve can be approximated by a function (Caughlan, 1988)

$$N_A \langle \sigma v \rangle = a_1 / T_9^{1/3} \exp(-a_2 / T_9^{2/3}) \cdot (1.0 + a_3 T_9^{1/3} + a_4 T_9^{2/3} + a_5 T_9 + a_6 T_9^{4/3} + \\ + a_7 T_9^{5/3} + a_8 T_9^{6/3} + a_9 T_9^{7/3}) + a_{10} / T_9^{a_{11}} \tag{4}$$

The parameters of such an approximation are given in Table 1.

Table 1. Parameterization parameters of (4) for reaction rate.

| № | $a_i$ |
|---|---|
| 1 | 3.078E-8 |
| 2 | 1.37107 |
| 3 | 1.53026E9 |
| 4 | -6.06099E9 |
| 5 | 3.10659E9 |
| 6 | 1.09445E10 |
| 7 | -2.07862E9 |
| 8 | -2.54614E10 |
| 9 | 1.96587E10 |
| 10 | -2.51227 |
| 11 | -6.04299 |

The result of the rate calculation using the (4) formula with such parameters is shown in Fig. 2 by dashed curve, which practically merges with continuous curve, with an average value of $\chi^2 = 0.6$. To calculate $\chi^2$, the error of the calculated data was assumed to be 5%.

**Conclusions.** The methods used in this work for obtaining the shape and depth of intercluster interaction potentials for scattering and bound states make it possible to get rid of the discrete and continuous ambiguities of their parameters. This solves a well-known problem that arises when constructing intercluster potentials in the continuous and discrete spectrum of a two-particle system by the usual optical method. The reaction rate was obtained and its approximation by a simple analytical expression was performed. Subsequently, the potentials obtained using these methods can be used in any calculations related to the solution of various nuclear-physical and astrophysical problems formulated at low, ultra-low and thermal energies.

**Acknowledgments.** The work was supported by the grant of the Ministry of Education and Science of the Republic of Kazakhstan № AP19676483 "Study of the processes of thermonuclear burning of hydrogen in the CNO cycle in the Sun and in stars" through the V.G. Fesenkov Astrophysical Institute ASC MDDIAI RK.